# Regular-Triangle Trimer and Charge Order Preserving the Anderson Condition in the Pyrochlore Structure of CsW$_2$O$_6$

Yoshihiko Okamoto[1*], Haruki Amano[1], Naoyuki Katayama[1], Hiroshi Sawa[1], Kenta Niki[1], Rikuto Mitoka[1], Hisatomo Harima[2], Takumi Hasegawa[3], Norio Ogita[3], Yu Tanaka[4], Masashi Takigawa[4], Yasunori Yokoyama[1], Kanji Takehana[5], Yasutaka Imanaka[5], Yuto Nakamura[1], Hideo Kishida[1], and Koshi Takenaka[1]

[1]*Department of Applied Physics, Nagoya University, Furo-cho, Chikusa-ku, Nagoya 464-8603, Japan.*

[2]*Department of Physics, Kobe University, Rokkodai 1-1, Nada-ku, Kobe 657-8501, Japan.*

[3]*Graduate School of Integrated Arts and Sciences, Hiroshima University, Kagamiyama 1-7-1, Higashi-Hiroshima, 739-8521, Japan.*

[4]*Institute for Solid State Physics, University of Tokyo, Kashiwanoha 5-1-5, Kashiwa 277-8581, Japan.*

[5]*National Institute for Materials Science (NIMS), Sakura 3-13, Tsukuba 305-0003, Japan.*

Since the discovery of the Verwey transition in magnetite, transition metal compounds with pyrochlore structures have been intensively studied as a platform for realizing remarkable electronic phase transitions. We report the discovery of a unique phase transition that preserves the cubic symmetry of the β-pyrochlore oxide CsW$_2$O$_6$, where each of W 5$d$ electrons are confined in regular-triangle W$_3$ trimers. This trimer formation is an unprecedented self-organization of $d$ electrons, which can be resolved into a charge order satisfying the Anderson condition in a nontrivial way, orbital order caused by the distortion of WO$_6$ octahedra, and the formation of a spin-singlet pair in a regular-triangle trimer. Electronic instability due to the unusual three-dimensional nesting of Fermi surfaces and the localized nature of the 5$d$ electrons characteristic of the pyrochlore oxides were found to play important roles in this unique charge-orbital-spin coupled phenomenon.

Understanding the electronic phase transitions of crystalline solids is a central issue in condensed matter physics. Metal–insulator transitions in transition metal compounds with pyrochlore structures, made of three-dimensional networks of corner-sharing tetrahedra, have posed challenging questions in materials science since their discovery. The classical example is magnetite Fe$_3$O$_4$, which shows a metal–insulator transition accompanied by a charge order of Fe at 119 K, called the Verwey transition[1]. Although many studies of this transition have been made, full understanding of its ground state has not yet been reached, and relevant studies based on new perspectives are continuing[2]. Recently, metal–insulator transitions accompanied by all-in-all-out-type magnetic order in 5$d$ oxides, such as Cd$_2$Os$_2$O$_7$ and Nd$_2$Ir$_2$O$_7$, have attracted considerable attention[3-5], in terms of a ferroic order of extended magnetic octapoles and the formation of Weyl fermions in solids[3,6-9]. As described above, rich physics appears in pyrochlore systems, which might be caused by the high crystal symmetry and a large number of atoms in a unit cell, resulting in the self-organization of $d$ electrons in various forms.

In this study, we report a unique self-organization of 5$d$ electrons at the metal–insulator transition of β-pyrochlore oxide CsW$_2$O$_6$, discovered by using high-quality single crystals. CsW$_2$O$_6$ was first synthesized by Cava *et al*., which was reported to have a cubic lattice with $Fd\bar{3}m$ space group at room temperature[10]. In this structure, W atoms form a pyrochlore structure and have 5.5+ valence with a 5$d^{0.5}$ electron configuration. Electrical resistivity measurement of polycrystalline samples suggested that a metal–insulator transition occurs at 210 K[11]. The crystal structure of the insulating phase was reported to have orthorhombic $Pnma$ space group[11]; however, this space group was suggested to be incorrect via electronic structure calculations[12]. Recent photoemission experiments of thin films suggested that the valence of W in the insulating phase disproportionates into 5+ and 6+[13].

Single crystals of CsW$_2$O$_6$ (Fig. 1A) were prepared in a quartz tube with a temperature gradient (see Method section).

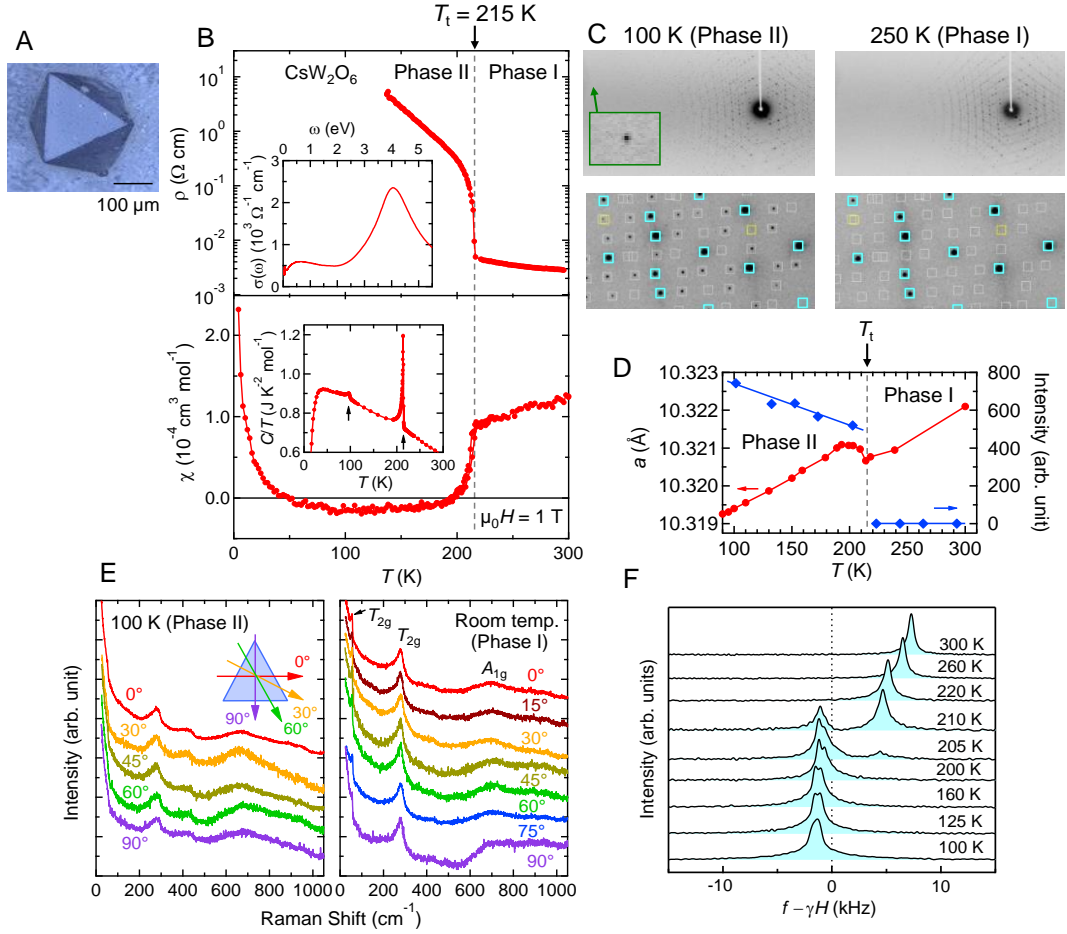

**Fig. 1. Physical and structural properties of $CsW_2O_6$.** All the experimental data in this figure except for the heat capacity were obtained using single crystals. **(A)** A single crystal of $CsW_2O_6$. **(B)** Temperature dependence of the electrical resistivity (upper) and magnetic susceptibility (lower). The inset of the upper panel shows the optical conductivity at room temperature deduced from the reflectivity using the Kramers-Kronig transformation[36]. The inset of the lower panel shows the heat capacity divided by temperature measured by using a polycrystalline sample. **(C)** Single-crystal XRD patterns at 250 (Phase I, right) and 100 K (Phase II, left). The pale blue and white squares indicate the positions of allowed and forbidden reflections, respectively, for the $Fd\bar{3}m$ space group. **(D)** Temperature dependence of the lattice constant and intensity of (−2, −5, −3) reflection estimated with the single-crystal XRD data. **(E)** Polarization dependence of the (111) Raman spectra measured at room temperature (Phase I, right) and 100 K (Phase II, left). **(F)** Temperature dependence of the $^{133}Cs$-NMR spectra measured in a magnetic field of 8 T applied along [001].

As shown in Fig. 1B, the electrical resistivity, $\rho$, of a single crystal strongly increases below $T_t$ = 215 K with decreasing temperature, as in the cases of a polycrystalline sample and a thin film[11,13]. This increase is accompanied by a small but obvious temperature hysteresis, indicating that a first-order phase transition occurs at $T_t$. Here, the phases above and below $T_t$ are named Phase I and II, respectively. The magnetic susceptibility, $\chi$, shown in Fig. 1B strongly decreases below $T_t$, which is identical to the polycrystalline case[11]. However, the line widths of the $^{133}Cs$-NMR spectra in Phase II do not show any significant broadening compared to those in Phase I, as shown in Fig 1F, indicating that the decrease of $\chi$ in Phase II is not caused by antiferromagnetic order.

Figure 1C shows single-crystal X-ray diffraction (XRD) patterns of $CsW_2O_6$ measured at 250 (Phase I) and 100 K (Phase II). Each of the diffraction spots at 250 K were indexed on the basis of a cubic cell of $a$ = 10.321023(7) Å with $Fd\bar{3}m$ space group, consistent with previous reports[10,11]. In the diffraction pattern at 100 K, more diffraction spots appear. All these spots were indexed on the basis of cubic $P2_13$ space group with a lattice constant of $a$ = 10.319398(6) Å, which is almost identical to $a$ of Phase I. This change of diffraction spots occurs at $T_t$, as seen in the temperature dependence of the intensity shown in Fig. 1D. Moreover, in Phase II, diffraction spots do not split into multiple spots nor do they change their shapes, even in the high-angle region, as shown in Fig. 1C. Laue class and crystal system checked by the observed reflections indicated that a structural change that preserves the cubic symmetry occurs at $T_t$. As seen in the polarization dependence of the Raman spectra of (111) sur-face measured at 100 K (Phase II) and room temperature (Phase I) shown in Fig. 1E, the spectra of Phase II are independent of the polarization angle same as in Phase I, indicating the presence of three-fold rotational symmetry perpendicular to (111), consistent with the inferred cubic symmetry. The crystallographic parameters at 250 and 100 K determined by the structural analyses are shown in Tables S2 and S4, respectively.

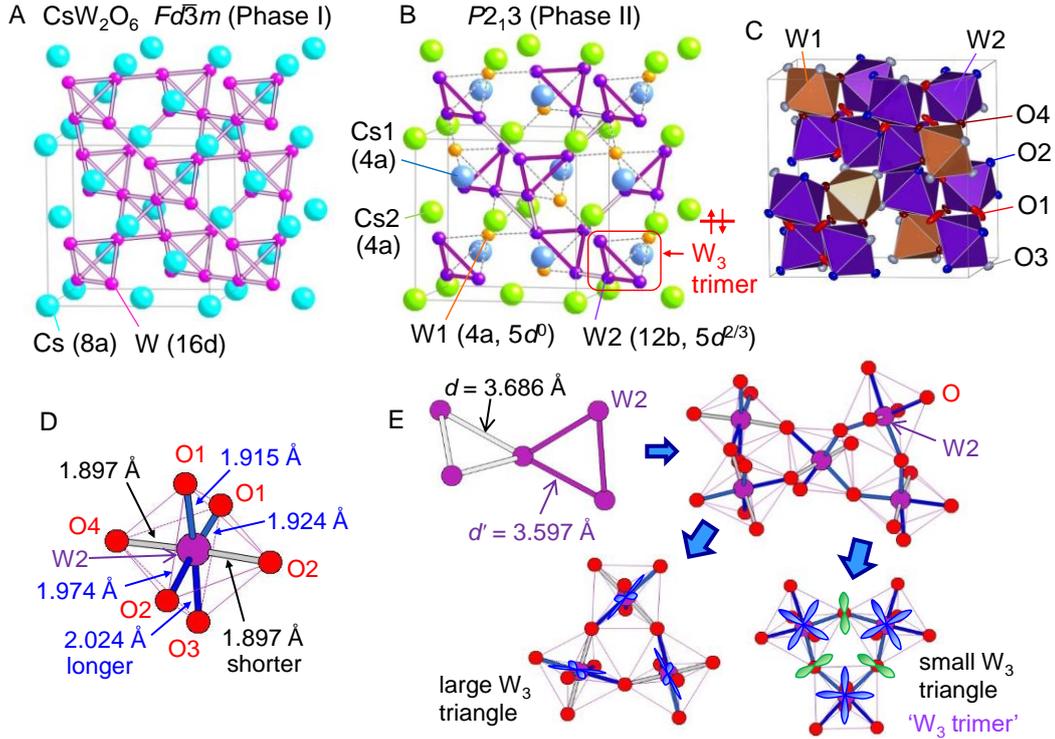

**Fig. 2. Crystal structure of CsW$_2$O$_6$.** (A,B) Crystal structures of the W and Cs sublattices of Phase I ($T > 215$ K) and II ($90 < T < 215$ K). The purple triangles indicate the W$_3$ trimers. (C) Thermal ellipsoids of the oxygen atoms at 100 K (Phase II). (D) W-O bond lengths in a W(2)O$_6$ octahedron at 100 K (Phase II). (E) Schematic picture of the occupied 5$d$ orbitals in Phase II. The lower left and right panels indicate the W$_3$O$_{15}$ units forming the larger and smaller triangles made of W atoms, respectively. The small triangle corresponds to the W$_3$ trimer. In the right panel, the 2$p$ orbitals of the bridging oxygen atoms are also shown.

Here we discuss the solved crystal structure of Phase II. In Phase I with the $Fd\bar{3}m$ space group, each of Cs, W, and O atoms occupies one site, where the Cs and W atoms form diamond and pyrochlore structures, respectively (Fig. 2A). In Phase II with the $P2_13$ space group, the Cs atoms occupy two different sites and form a zinc-blende structure, as shown in Fig. 2B. This was further confirmed by the two peaks in $^{133}$Cs-NMR spectra correspond to the two Cs sites, which appear as a small peak split in the 200, 160, and 125 K data shown in Fig. 1F. On the other hand, W atoms occupy two sites with a 1:3 ratio in Phase II, as shown in Fig. 2B, which is incompatible with the W$^{5+}$-W$^{6+}$ charge order with a 1:1 ratio of W$^{5+}$ and W$^{6+}$ atoms. According to the bond valence sum calculation for the W-O distances determined from single-crystal XRD analyses[14], the valences of the W(1) and W(2) atoms are estimated to be 6.07 and 5.79 at 100 K (Phase II), respectively. Considering that the reliable bond valence sum parameters of W$^{6+}$ are available but those of W$^{5+}$ are not, it is natural for the W(1) atoms to be W$^{6+}$ without 5$d$ electrons. In this case, the valence of the W(2) atoms become 5.33+ with 5$d^{2/3}$ electron configurations. The above discussion indicates that a charge order with a noninteger valence occurs at $T_t$. In fact, W-deficient CsW$_{1.835}$O$_6$ crystals, where all W atoms have 6+ valence without 5$d$ electrons, do not show the transition at $T_t$ (see Supplementary Information).

In Phase II, the W(2) atoms form a three-dimensional network of small and large regular triangles, which are alternately connected by sharing their corners, as shown in Fig. 2B. Although the difference of sizes between the large ($d$) and small triangles ($d'$) are about 2%, arrangements of the occupied 5$d$ orbitals are completely different between them, resulting in a W$_3$ trimer on a small triangle, as discussed later. If there was no alternation of the W$_3$ triangles, the W sublattice would possess a hyperkagome structure[15]. The presence of the alternation indicates that 'breathing hyperkagome' structure is formed during Phase II[16].

This charge order is interesting in that the 'Anderson condition' is maintained in a unique way. P. W. Anderson pointed out that magnetite has an infinite number of charge ordering patterns, where all the tetrahedra in a pyrochlore structure have the same total charge, i.e., so-called Anderson condition, and this macroscopic degeneracy strongly suppresses the transition temperature of the Verwey transition[17]. This situation can be interpreted as geometrical frustration of electronic charges. However, not only magnetite, but also other mixed-valent pyrochlore systems, such as $CuIr_2S_4$ and $AlV_2O_4$, were reported to show a charge order that violated the Anderson condition[18-21]. In them, the energy gained by σ bonding between $d$ orbitals of adjacent atoms was expected to be large enough to compensate for the loss of Coulomb energy due to violation of the Anderson condition, because spinel-type compounds comprised the edge-shared octahedra[22,23]. In contrast, the charge order of $CsW_2O_6$ satisfies the Anderson condition, where each tetrahedra consists of three $W^{5.33+}$ and one $W^{6+}$ atoms. However, this charge order is different to that proposed by Anderson and Verwey, which has integer valences with a 1:1 ratio[17,24]. Hyperkagome-type orders often appear in pyrochlore systems with a 1:3 ratio of two kinds of atoms, such as the uuud spin structure of the half magnetization plateau of Cr spinel oxides and the atomic order in B-site ordered spinel oxides $A_2BB'_3O_8$[25-27]. As far as we are aware, $CsW_2O_6$ is the first example to show a hyperkagome-type order where the formation of this order is nontrivial. It is a novel way to relieve the geometrical frustration based on the traditional problem in condensed matter physics.

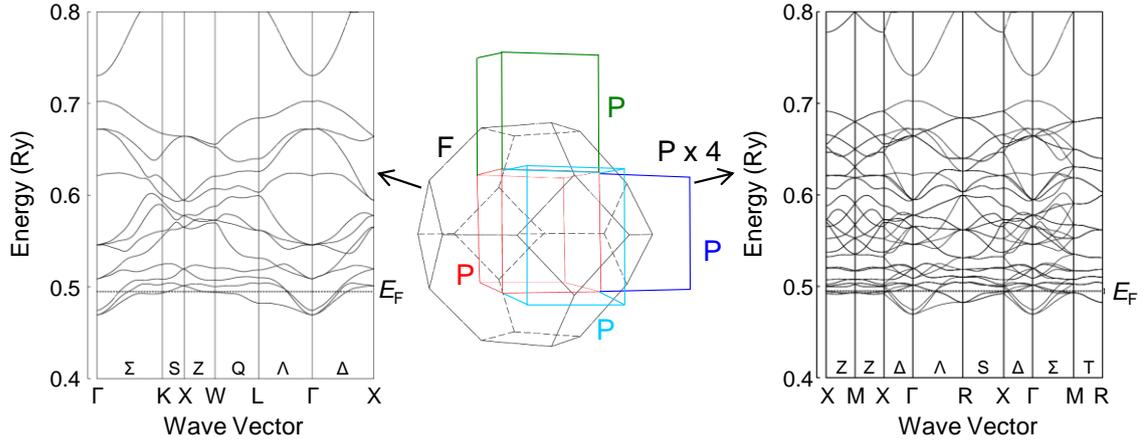

**Fig. 3. First principles calculations of $CsW_2O_6$ of Phase I calculated with spin-orbit coupling.** The left panel is the electronic band structure based on the Brillouin zone of the face-centered cubic lattice, which is shown by black solid lines in the center panel. In the right panel, four band structures based on four primitive cells, which are shown by red, green, pale blue, and blue solid lines in the center panel, respectively, overlap in the graph.

Why does such a unique charge order occur in $CsW_2O_6$? A key to understand this question is hidden in Fermi-surface instability of the electronic band structure of Phase I. Figure 3A shows the band structure of Phase I and Figure 3B shows four overlapping band structures, which are depicted after the parallel shifts of electronic bands corresponding to a change of the primitive cell from face-centered cubic to primitive cubic. As seen in Fig. 3B, band crossing occurs close to all points where electronic bands touch the Fermi energy $E_F$, indicating that the Fermi surfaces are well nested by the parallel shift of the electronic bands, corresponding to the loss of centering operations. This situation can be called 'three-dimensional nesting', which suggests that a large electronic energy is gained by the structural distortion associated with the above symmetry change. It is quite rare for cubic compounds to have such well-nested Fermi surfaces, except for the filled-skutterudite $PrRu_4P_{12}$. $PrRu_4P_{12}$ shows a metal−insulator transition accompanied by a structural change from body-centered cubic to primitive cubic[28,29] and has a Fermi-surface instability corresponding to this structural change[30].

The above discussion clearly indicates that three-dimensional nesting is an essential ingredient for the 215 K transition. However, if it is the only driving force, a structural change from $Fd\bar{3}m$ to $P4_132$ or $P4_332$, which are maximal non-isomorphic subgroups with primitive cubic lattices, must occur. In this case, the W(2) atoms should form a uniform hyperkagome structure. In reality, the space group of Phase II is $P2_13$, which is a subgroup of $P4_132$ and $P4_332$, and $d$ is 2% shorter than $d'$. Phenomenologically, orientation of the occupied 5$d$ orbitals is important for this symmetry lowering. For a $W(2)O_6$ octahedron of Phase II shown in Fig. 2D, the two apical W(2)-O bonds (gray) are 3−8% shorter than the other four equatorial bonds (blue), indicating that the octahedron is uniaxially compressed. This compression is comparable to the typical Jahn-Teller distortion in $t_{2g}$ electron systems, and the 5$d$ orbitals lying in this equatorial plane should be occupied by electrons. Schematic pictures of the occupied 5$d$ orbitals in small and large triangles are shown in the right and left panels of Fig. 2E, respectively. There is considerable overlap between the occupied 5$d$ orbitals in the small triangle via an O 2$p$ orbital. In contrast, there is little overlap in a large triangle, indicating that two electrons in three W(2) atoms are confined in a W$_3$ trimer in the small triangle.

This regular-triangle trimer formation might be understood as a three-centered-two-electron (3c2e) bond formation, where two electrons are accommodated in a molecular orbital made of three W 5$d$ orbitals (and the O 2$p$ orbitals hybridize with them). In this case, it is natural to have a nonmagnetic ground state. This regular-triangle trimer is essentially different to those of famous LiVO$_2$ and LiVS$_2$, where two electrons are shared by two V atoms along each side of a triangle[31,32]. Stabilization of the electronic energy by the formation of multiple-centered bonds, where a few electrons are shared by many atoms, often occurs in electron-deficient molecules or cluster compounds[33]. To our knowledge, CsW$_2$O$_6$ is the first example where this type of bond formation appears as a phase transition. The formation of regular-triangle trimers itself is also surprising, because the 3c2e bond usually has a bent shape. According to previous reports, only the H$_3^+$ ion has a regular-triangle shape. Moreover, H$_3^+$ is an interstellar material and it is not stable on Earth, having been observed in astronomical spectra[34].

This unique regular-triangle shape of the trimer might be related to its internal structure, a part of which appears in the atomic displacement parameters (ADPs). As shown in Fig. 2C, the O atoms bridging W(2) atoms in a W$_3$ trimer (O(1) site) in Phase II have large ADPs perpendicular to the W-W bond. The ADPs of the other O atoms are typical values, suggesting that the ADPs of the O(1) site do not increase by the structural instability of the β-pyrochlore structure, but rather by the electronic instability of the trimer. The large ADPs perpendicular to the W-W bond indicate that there is a strong fluctuation that changes the W(2)-O(1)-W(2) angle. In pyrochlore oxides, the change of this angle has a large effect on the orbital overlap[35]. Therefore, this fluctuation can be interpreted as a strong fluctuation to a state in which one of the W-W bonds becomes stronger, or in the extreme, to a state in which a W$_2$ dimer is formed. Since ADPs of the W(2) site have typical values, it is unlikely that the dimers statically and randomly form on the trimers. Instead, the dimer might dynamically fluctuate or resonate. For a complete understanding of the internal structure of the trimers, it would be desirable to directly observe their dynamical properties in a future study.

Another essential factor for the formation of this unique trimer is the localized nature of the 5$d$ electrons in CsW$_2$O$_6$. The optical conductivity spectra of CsW$_2$O$_6$ measured at room temperature deduced from the reflectivity using the Kramers-Kronig transformation[36], shown in the inset of Fig. 1B, exhibit a broad peak at around 0.6 eV. Extrapolation of the spectra to zero frequency coincides with ρ = 3 mΩ cm at room temperature (the main panel of Fig. 1B). This result indicates that there is no, or negligibly small, Drude contribution in the spectra, and the conducting carriers are trapped by something with an energy scale of 0.6 eV, resulting in the loss of coherency, which is supported by $d\rho/dT < 0$ in Phase I. Absence of a peak in the far-infrared region indicates that this localization is not due to disorder, but is reminiscent of the spectra of lightly-carrier-doped Mott insulators[37-39]. As seen in the band structure shown in Fig. 3A, there are flat parts near $E_\text{F}$ and the energy bands have a narrow width of 0.7 eV, suggesting the presence of a strong electron correlation for a 5$d$ electron system. In 5$d$ or 4$d$ pyrochlore oxides, the 5$d$/4$d$ electrons often have localized natures because of the small orbital overlap due to the bent metal-oxygen-metal bonds. In fact, the optical conductivities of Nd$_2$Ir$_2$O$_7$ and Sm$_2$Mo$_2$O$_7$ indicate the presence of incoherent $d$ electrons, similar to the case of CsW$_2$O$_6$[9,40]. As a result, 5$d$ pyrochlore oxides often show an electronic phase transition with the order of electronic degrees of freedom. Nd$_2$Ir$_2$O$_7$ with $J_\text{eff}$ = 1/2 and Cd$_2$Os$_2$O$_7$ with $S$ = 3/2, without charge and orbital degrees of freedom, showed a magnetic order accompanied by a metal–insulator transition[3-5]. Instead, for CsW$_2$O$_6$, the trimers are formed by the charge and orbital order, in which the two 5$d$ electrons form a spin-singlet pair, resulting in the nonmagnetic and insulating ground state. This is a novel type of self-organization of $d$ electrons realized in a strongly correlated 5$d$ oxide.

Finally, we will discuss another structural transition at 90 K. Phase II looks like a ground state, where most of the degrees of freedom have been lost, but surprisingly another phase transition occurs at 90 K. By indexing the diffraction spots in the single-crystal XRD data of Phase II, the crystal structure below 90 K, named Phase III, was found to have monoclinic $P2_1$ space group with a four-times-larger (2 × 1 × 2) unit cell than that of Phase II, as shown in Fig. S1. As seen in the inset of Fig. 1B, the heat capacity divided by temperature, $C/T$, shows a small but obvious peak at approximately 90 K, indicating the presence of a bulk phase transition. The atomic positions in Phase III have not yet been determined, because of tiny monoclinic distortion and domain formation, but it is clear that the structural change at 90 K is small, as seen in Fig. S2. In addition, χ does not exhibit an anomaly at 90 K. These results strongly suggest that the 90 K transition is not caused by the spin, charge, and/or orbital order different to the 215 K transition.

What mechanism gives rise to the 90 K transition? The diffuse scattering that appears in the single-crystal XRD patterns might provide a hint to answering this question. In the single-crystal XRD patterns of Phases I, II, and III shown in Fig. S4A, there are diffuse scatterings at the same positions, which follow the extinction rule of $h + l = 4n$ (for a cubic unit cell) and connect the superlattice spots that emerged in Phase III. This suggests that the structural change from Phase II to III and the diffuse scatterings have the same origin. The same diffuse scattering pattern also appeared in CsW$_{1.835}$O$_6$ (Fig.

S5) and CsTi$_{0.5}$W$_{1.5}$O$_6$[41], which are isostructural to CsW$_2$O$_6$, but only have W$^{6+}$ atoms without 5$d$ electrons, suggesting that they are independent of the 215 K transition and might be caused by the structural instability of the β-pyrochlore structure itself. This discussion also implies that the 215 K transition is irrelevant to this instability and is purely electronic driven.

In conclusion, we found that regular-triangle W$_3$ trimers are formed at the 215 K transition in β-pyrochlore oxide CsW$_2$O$_6$, as determined using structural- and electronic-property measurements of high-quality single crystals. This transition represents an unprecedented self-organization of 5$d$ electrons, where geometrical frustration is relieved in a nontrivial way that satisfies the traditional Anderson condition and results in the quite rare cubic–cubic structural transition. This type of electronic transition is not only unique, but is only partly understood by our first principles calculations, suggesting that it might be a spin-, charge-, and orbital-coupled phase transition occurring beyond the existing electronic phase transitions of pyrochlore systems. The above finding shows that the exploration of novel geometrically frustrated 5$d$ compounds will lead to the discovery of further new electronic phenomena, such as odd-parity multipoles and spin-charge-orbital entangled quantum liquids.

**Methods**

Single crystals of CsW$_2$O$_6$ were prepared by crystal growth in an evacuated quartz tube under a temperature gradient. A mixture of a 3:1:3 molar ratio of Cs$_2$WO$_4$ (Alfa Aeser, 99.9%), WO$_3$ (Kojundo Chemical Laboratory, 99.99%), and WO$_2$ (Kojundo Chemical Laboratory, 99.99%), with a combined mass of 0.1 g, was sealed in an evacuated quartz tube with 0.1 g of CsCl (Wako Pure Chemical Corporation, 99.9%). The hot and cold sides of the tube were heated to, and then kept at, 973 and 873 K, respectively, for 96 h, and then the furnace was cooled to room temperature. The mixture was put on the hot side. The obtained single crystals had an octahedral shape with {111} faces with edges of at most 1 mm. Powder samples of CsW$_2$O$_6$ were prepared by the solid-state reaction method described in Refs. 10 and 11. The obtained powder was sintered at 773 K for 10 min using a spark plasma sintering furnace (SPS Syntex).

Single crystals of W-deficient CsW$_{2-x}$O$_6$ were prepared using the flux method. A mixture of a 3:1:3 molar ratio of Cs$_2$WO$_4$ (Alfa Aeser, 99.9%), WO$_3$ (Kojundo Chemical Laboratory, 99.99%), and WO$_2$ (Kojundo Chemical Laboratory, 99.99%), with the combined mass of 0.1 g, and 0.2 g of CsCl (Wako Pure Chemical Corporation, 99.9%) were put in an alumina crucible, which was sealed in an evacuated quartz tube. The tube was heated to, and then kept at, 923 K for 48 h, and then slowly cooled to 873 K at a rate of −0.5 K/h. The obtained single crystals have similar octahedral shape and mostly have a larger size than those of CsW$_2$O$_6$. The value of the W deficiency, $x$, was estimated to be 0.165 via a structural analysis using the single-crystal X-ray diffraction (XRD) data, meaning that the chemical composition of the single crystal is CsW$_{1.835}$O$_6$, where the W atoms have no 5$d$ electrons, as mentioned later.

The electrical resistivity and magnetization measurements of the CsW$_2$O$_6$ single crystals were performed using a Physical Property Measurement System (PPMS, Quantum Design) and Magnetic Property Measurement System (Quantum Design), respectively. The normal incident reflectivity of (111) surface of a CsW$_2$O$_6$ single crystal was taken at room temperature using a Fourier-type interferometer (0.005–1.6 eV, DA-8, ABB Bomem; 0.06–1.0 eV, FT-IR6100, Jasco) and a grating spectrometer (0.46–5.8 eV, MSV-5200, Jasco) installed with a microscope[36,42]. As a reference mirror, we used either evaporated Au (far- to near-IR region), Ag (near-IR to visible region), or Al (near- to far-UV region) films on a glass plate. The heat capacity of the CsW$_2$O$_6$ sintered sample was measured using the relaxation method with the PPMS. Single-crystal XRD experiments of the CsW$_2$O$_6$ and CsW$_{1.835}$O$_6$ samples were performed at BL02B1 in the SPring-8 synchrotron radiation facility in Japan. The experimental conditions are shown in Tables S1, S3, and S5. SORTAV and SHELXL were used for merging the reflection data and the structural refinement[43-45]. A part of crystal structure views were drawn using VESTA[46]. Powder XRD experiments of CsW$_2$O$_6$ were performed at BL02B2 in SPring-8. Synchrotron X-rays with energies of 15.5 and 25 keV were used for the measurements below and above 150 K, respectively. Rietveld analyses of the powder XRD data were performed using GSAS. Raman scattering spectra of the CsW$_2$O$_6$ single crystals were measured using a diode-pumped CW solid-state laser with a wavelength of 5614 Å. $^{133}$Cs-NMR measurements of a CsW$_2$O$_6$ single crystal were conducted in a magnetic field of 8 T. The NMR spectra were obtained by Fourier transforming the free induction decay signal. The band structure calculations of Phase I of CsW$_2$O$_6$ were performed using the full potential linear augmented plane wave (FLAPW) method with a local density approximation. The experimentally obtained structural parameters were used for the calculations.

**Acknowledgments**

The authors are grateful to Y. Yamakawa, A. Yamakage, A. Koda, R. Kadono, J. Matsuno, and D. Hirai for helpful discussions and T. Fujii for his help with the experiments. Y. O. is also grateful for collaboration with Y. Nagao, J. Yamaura, M. Ichihara, Z. Hiroi, and M. Yoshida in the early stage of this work using polycrystalline samples. This work was partly carried out under the Visiting Researcher Program of the Institute for Solid State Physics, the University of Tokyo and supported by JSPS KAKENHI (Grant Number: 18H04314, 19H05823, 16H03848, 15H05886, 15H05882, 20K03829, 17H02918, 18H04310).

*E-mail: yokamoto@nuap.nagoya-u.ac.jp